# Observation of superconducting phase transition in InN

Zhi-Yong Song,[1,2] Liyan Shang,[1] JunHao Chu,[1,2,*] Ping-Ping Chen,[2] Akio Yamamoto[3] and Ting-Ting Kang[2,*]

[1]*Key Laboratory of Polar Materials and Devices, Ministry of Education, East China Normal University, Shanghai 200062, China*
[2]*State Key Laboratory of Infrared Physics, Shanghai Institute of Technical Physics, Chinese Academy of Sciences, 200083 Shanghai, People's Republic of China*
[3]*University of Fukui, Fukui 910-8507, Japan*

* E-mail: jhchu@mail.sitp.ac.cn (J. C.)
*E-mail: kang@mail.sitp.ac.cn (T. K.)

**Abstract:** InN superconductivity is very special among III-V semiconductors, because other III-V semiconductor (like GaAs, GaN, InP, InAs etc.) usually lacks strong covalent bonding and seldom shows superconductivity at low-temperature. In this paper, via current-voltage(I-V) measurement, we probe the superconducting phase transitions in InN. The possible connection with those chemical-unstable phase separated inclusions, like metallic indium or $In_2O_3$, was removed by HCl acid etching. It finds InN samples can show different phase transition behaviors. The vortex-glass (VG) to liquid transition, which is typical in type-II superconductors, is observed in the sample with large InN grain size. In contrast, the small grain-sized sample's superconducting properties are sensitive to acid etching, shows a transition into a non-zero resistance state at the limit of temperature approaches zero. Our work suggests that the grain size and inter-grain coupling may be two key factors for realizing InN superconductivity. InN superconductivity can become robust and chemical stable if the grain size and inter-grain coupling both are large enough.

## I. INTRODUCTION

Group-III nitrides are the typical semiconductors interested for optoelectronic applications [1-3]. Therefore, it is surprising that one of its members – InN [4,5] can be superconductor. However, upon its first report by Inushima et al.[6], InN superconductivity and its importance had not been understood by most people, yet. It is known that typical III-V semiconductor (e.g. GaAs, InAs, GaN etc., except BN, whose B and N atoms are bound by strong covalent bonds) is usually not easy to be tuned into superconducting [7,8]. In contrast, IV semiconductor (Si, diamond, SiC etc.) can become superconductor by heavy doping [9]. The reason is explained under the framework of BCS mechanism: (1) IV semiconductor is with strong directional covalent bonds, but III-V semiconductor is with weak covalent bonds. (2) III-V semiconductor is mostly made by heavy element. Therefore, III-V semiconductor doesn't have large enough phonon energy and a strong electron-phonon coupling. Additionally, the enhanced spin- orbit coupling in III-V tends to break the cooper pairing under BCS mechanism. In this sense, even III-nitride can be superconducting, GaN(which is with more covalent character and made by lighter element) shall be easier to be superconductor than InN. However, the experimental evidence on GaN superconductivity is still lack [7]. The traditional IV semiconductor superconductivity mechanism, which is driven by phonons strongly coupled to holes at the Γ point and requires the material to be p-type [10], is not straightforwardly applicable in InN. Because InN superconductivity is all observed in n-type InN up to now[11-13].

Although InN superconductivity seems to be "unreasonable" within the framework of current superconductivity theory for semiconductors, the major doubt on InN superconductivity is rooted in a simple speculation: InN can have two phase separated superconducting inclusions, i.e. indium metal (In) and indium oxide($In_2O_3$), whose superconductivity may "contaminate" InN. However, this naive $In/In_2O_3$ speculation faces strong challenges. Concerning In, its upper critical field $H_{c2}$ (~0.03T) is much smaller than that of InN(~1T). Although the small sized In nanoparticle had been theoretically proposed to increase $H_{c2}$ [14], such proposal is not reliable because small size effect had been experimentally proved to suppress the superconductivity in In [15]. Furthermore, another similar III-nitride – GaN had been tuned into superconducting by containing a heavy amount of Ga and has a similar $H_c$ as Ga [7,16]. Then it is less persuasive to assume that metal nanoparticle superconductivity effect is only present in InN. On $In_2O_3$, the attribution of 33°XRD peak to $In_2O_3$ needs further clarification [17], because metallic indium, amorphous InN, InN(10-11), cubic InN can also produce similar XRD reflection [12]. Additionally, achieving $In_2O_3$ superconductivity is not easy and needs low disorder [18]. And it is hard to believe the unintentionally introduced $In_2O_3$ in InN can meet such requirement.

Besides the importance in superconductivity physics, InN superconductivity is of technical interests. For example, InN superconductor can be used to fabricate the superconducting single photon detector (SSPD)[19,20]. Firstly, compared with NbN - the most successful material in SSPD, InN shares many similar and attractive properties, namely, high chemical stableness, good mechanical performance etc. Secondly, InN has a suitable transition temperature $T_c$(~3K). While high-$T_c$ superconducting material is not suitable for SSPD application, because its large superconducting gap energy will reduce the sensitivity to photons of a given energy (especially at longer wavelength) [20]. On the other hand, too low $T_c$ requires a large cooling ability and is less convenient. Finally, InN's application in superconducting industry is favored because it can be integrated with the III-V technology, allowing an easy combination of semiconductor and superconductor function within one material system, i.e. III-nitride.

In this work, we note that, if InN superconductivity is produced by $In/In_2O_3$, such superconductivity will be chemical unstable, because In/$In_2O_3$ is highly reactive. Therefore, it is straightforward to remove such inclusions by acid etching and check InN's superconductivity. And such experiment is not available in the literatures, yet. This work is also useful for the future applications of InN in SSPD, where a chemical stableness is preferred.

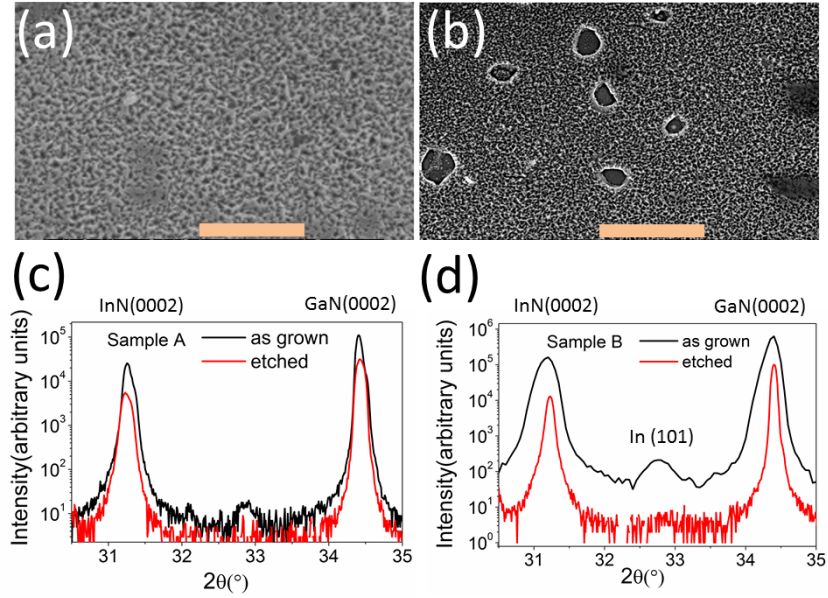

Fig.1. (a) SEM images for the etched sample A(a) and B(b), the scale bar's length is 5μm. (c,d) X-ray diffraction patterns of the as-grown, etched InN films of sample A(c) and B(d).

## II. EXPERIMENTAL METHOD

For comparison, we use two n-type InN films in this work, which are termed as sample A and B, respectively. Both of them were grown by metalorganic vapor phase epitaxy (MOVPE) on insulative GaN/sapphire (0001) templates in University of Fukui. The growth temperature is 520°C(480°C) for sample A(B). The detailed growth conditions can be found elsewhere [21]. The thickness of sample A(B) is measured to be ~850 nm (~950 nm) by cross-section SEM. The room temperature electron mobility and concentration of sample A(B) is 505$cm^2$/Vs(280$cm^2$/Vs) and $1.2×10^{19}cm^{-3}$($2.2×10^{19}cm^{-3}$). Fig.1 (c) and (d) are the X-ray diffraction (XRD) results of sample A and B, respectively. In addition to the InN(0002) reflection at 31.36°, both two samples have a small peak at ~33.0°. This small peak can be removed by the HCl acid etching. Therefore, it is attributed to In(101) (2θ=32.95°) [22], or (110) reflection of a rhombohedral phase $In_2O_3$(2θ=32.92°).

Each sample was cut into one small piece (sized ~1mm×10mm) and subjected to HCl etching to remove the possible In/In$_2$O$_3$ inclusions. Figure 1(a, b) shows the surface morphologies (SEM images) of the HCl etched samples. As seen, sample B shows more voids after HCL etching, indicating more serious phase separation. Transport measurements were done in Shanghai Institute of Technical Physics using a dilution refrigerator down to ~30mK temperature and a cryogen-free low-temperature system with lowest temperature of 1.5K. Magnetic field was applied vertical to the sample surface, i.e. along InN c[0001] axis. The Ohmic contacts were made using metal indium or silver pastes.

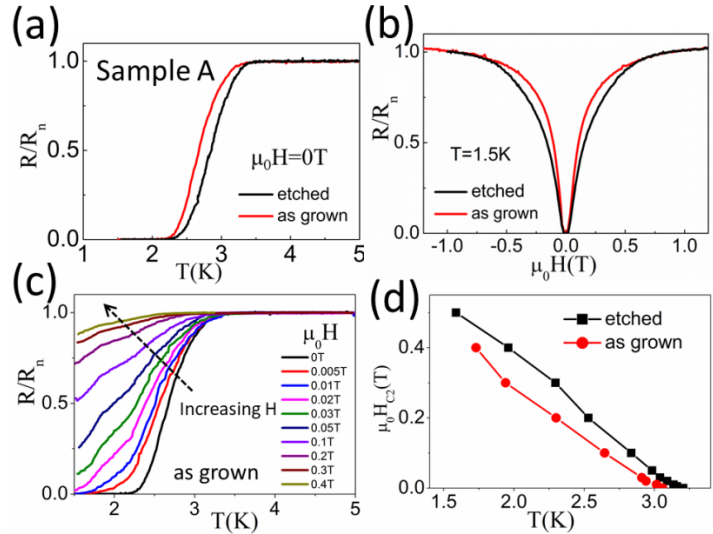

Fig.2. (a) R-T transitions and (b) R-B transitions for as-grown and etched sample A. (c) Normalized resistance as a function of temperature under different magnetic field for as-grown sample A. (d) The upper critical field ($H_{c2}$) as a function of temperature (T) for as-grown and etched sample A. In (a,b,c), the resistance is measured with a current of 1μA.

### III. RESULTS AND DISCUSSIONS

*3.1. The existence of stable InN superconductivity against chemical etching.*

In Fig.2(a) and (b), for sample A, the normalized resistance R/R$_n$ (R$_n$ is the normal state resistance) as a function of temperature T(R-T) and magnetic field H(R-H) is displayed. Because superconductivity is both observed in the as-grown and etched sample A, it proves that the superconductivity in InN can survive from the acid etching. And the critical temperature and magnetic field are not changed significantly by etching. These observation indicate those chemical stable inclusions, like large sized In/In$_2$O$_3$, is less possible to be responsible for InN's superconductivity in sample A.

Fig.2(c) displays the R-T curves of the as-grown sample A under different magnetic field. On increasing H, the superconducting transition broadens and shifts to lower temperature. Using these curves, we can deduce the upper critical field (H$_{c2}$), and H$_{c2}$ is defined as the H field where R/R$_n$=90%. In Fig.2 (d), we calculated the dirty-limit coherence length ($\xi_{0,dirty}$) of sample A by dirty limit(in case of the mean free path $l<\xi_0$) relation [23]:

$$H_{c2}(0) = 0.69 T_c \frac{dH_{c2}}{dT}|_{T=T_c} \quad (1a)$$
$$\xi_{0,dirty} = [\phi_0/2\pi H_{c2}(0)]^{1/2} \quad (1b)$$

where T$_{c2}$(H) at different H is determined from R-T scans where R/R$_n$=90% is met. Eq.(1) gives the similar $\xi_{0,dirty}$ value for the as grown(≈23nm) and etched(≈21nm) sample A. The clean-limit (Pippard) coherence length $\xi_{0,clean}$ of sample A is estimated by [24,25,26]:

$$\xi_{0,clean} = \frac{V_F \hbar}{\pi \Delta} = \frac{0.1804 V_F \hbar}{k_B T_C} \quad (2)$$

where 2Δ=E$_g$ is the superconducting energy gap, V$_F$ is the fermi velocity, T$_c$=2.4K is the transition temperature where R/R$_n$=10% happens. It reaches $\xi_{0,clean}$≈710nm.

In a conventional superconductor, coherence length ξ(T) will approach $\xi_0$ in the limit of T→0. While near T$_c$, ξ(T) diverges as (T$_c$-T)$^{-1/2}$. $\xi_0$ is equal to $\xi_{0,clean}$ if the sample is "clean" and to $\xi_{0,dirty}$ if it is "dirty" (i.e., when the mean free path $l<\xi_0$) [26]. By T>T$_c$ Hall measurements and InN's effective electron mass $m_e^*/m_0$=0.07 [5], we reach $l$≈25nm for sample A. Therefore, the dirty limit requirement $l<\xi_{0,dirty}$ is not met well and $\xi_0$ will be close to $\xi_{0,clean}$. $\xi_0$ is the low bound for the grain size which permits the occurrence of superconducting ordering [15,25] and we will address this topic later.

*3.2. The vortex liquid -glass transition.*

For the etched sample A, Fig.3(a) is the I-V(current-voltage) curves under different temperature, with a small magnetic field μ$_0$H=5mT (for introducing the vortex). Fig.3(c) displays the I-V curves under different magnetic field H, at a fixed temperature 0.5K(<<T$_c$). It shows that the I-V curves show a smooth evolution in the curvature from being convex to linearity with increasing temperature and magnetic field. These phenomena agree well with the vortex-glass (VG) theory, where the vortex comes from self-field due to persistent current and applied magnetic field is pinned/depinned. In VG theory, with the decrease of temperature, the mobile vortex (i.e. vortex liquid state) will finally be pinned by the randomly distributed pinning centers, reaching the superconducting state (i.e. vortex glass state).

According to VG theory, the I-V curves at different temperatures near liquid-glass phase transition temperature (T$_g$) can be scaled into two different branches by the scaling law [27,28,29]:

$$\frac{V}{I(T-T_g)^{\nu(z+2-D)}} = f_\pm \left(\frac{I}{|T-T_g|^{\nu(D-1)}}\right) \quad (3)$$

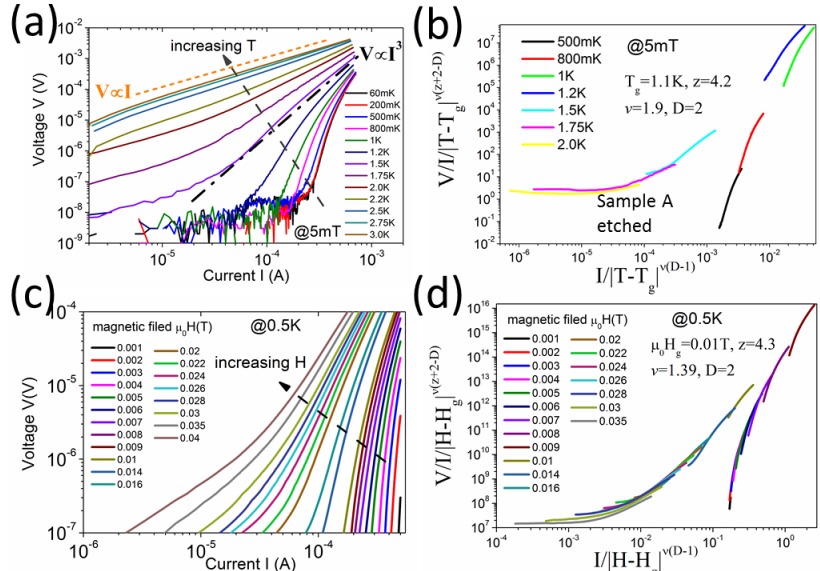

Fig.3. For etched sample A, (a) I-V curves at various temperatures from 35mK to 2.5K for μ$_0$H=5mT. (b) VG scaling of the V-I curves in (a). (c) I -V characteristics measured at 0.5 K for magnetic fields ranging from 0 to 40mT. (d)VG scaling of the V-I curves in (c).

where $v$ is the exponent of the vortex-glass correlation length $\xi_g$ and $\xi_g$ diverges at $T_g$ as:
$$\xi_g \sim |T-T_g|^v \quad (4)$$
$z$ is the dynamical critical exponent, D is the dimension number, and $f_\pm$ are the scaling functions above and below $T_g$. Above $T_g$, where in the vortex liquid phase, there is a finite linear resistance for low current limit $I \to 0$ [27]:
$$R_{lin} = (V/I)_{I\to 0} \propto (T-T_g)^{v(z+2-D)} \quad (5)$$
and the I-V curve goes non-linearly for large I. At $T_g$, the I-V curve satisfies the relationship:
$$V(I)|_{T=T_g} \sim I^{z+1/D-1} \quad (6)$$

Below $T_g$, where in the vortex-glass regime, the double-logarithmic I-V curves have downward curvatures corresponding to vanishing linear resistance. The scaling results are shown in Fig.3(b). The I-V curves are nicely scaled into two branches, which touches each other at the T=1K-1.2K scaling curves. This agrees well with the fitting $T_g$ value are separated into two parts by a particular magnetic field ($\mu_0 H_g$). We can modify Eq. (3) by replacing $T_g$ with $H_g$, based on VG theory [27]. Then those IV curves can also be scaled into two branches with $\mu_0 H_g$=10mT, D=2.

D=2 means the VG transition in Fig.3 is quasi-2D [29,30]. This is further supported by two behaviors in I-V. Firstly, as shown in Fig.3(a), for temperature in 1.5K-2K (i.e. the vortex liquid region), the I-V curves exhibit a crossover from linear behavior at small current I to nonlinear behavior at high I. Such a crossover is anticipated in the quasi-2D VG model: at low I, the vortex dynamics at length scales larger than $\xi_g$ ($\xi_g$ is of the order of $\xi$) is concerned, where the system will look like a vortex liquid with a linear resistance); at high I, the excitations involve length scales smaller than $\xi_g$ and the associated glassy dynamics yield a nonlinear I-V relation [29]. Secondly, a Berezinskii-Kosterlitz-Thouless (BKT) transition [30,31,32], which is a characteristic phase transition in 2D superconductor, was found at a specific temperature $T_{BKT}$=1.5K in Fig.3(a) by

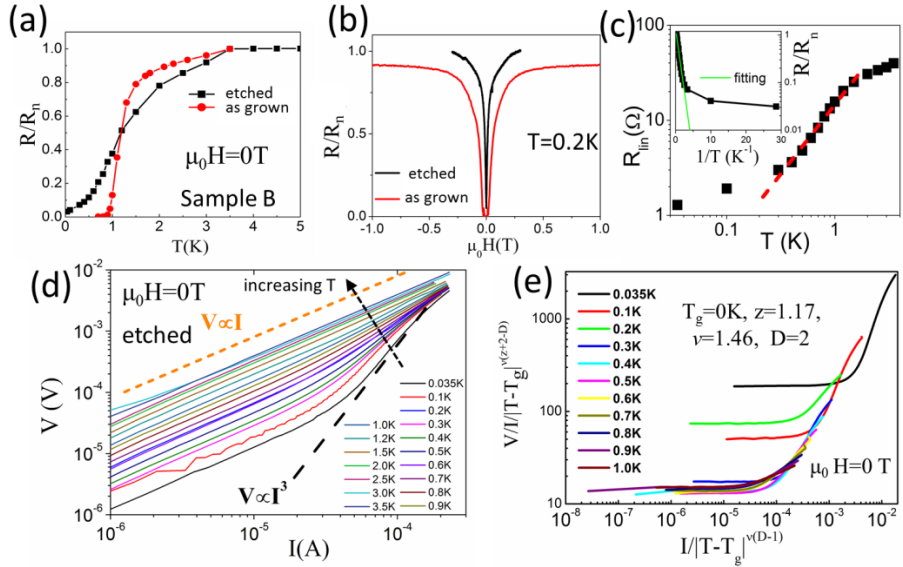

Fig.4. (a) R-T transitions and (b) R-B transitions for as-grown and etched sample B. (c) The linear resistance $R_{lin}$ at $I\to 0$ [deduced from the I-V curves in Fig.4(d)] as a function of temperature for etched sample B. (d) I-V characteristics measured at magnetic field $\mu_0 H$=0T for etched sample B. (e) Scaling of the V-I curves in (d).

$T_g$=1.1K.

Fig.3(a) can be fully explained within the framework of VG theory. When temperature $T<T_g$, the sample is in the vortex glass state. So the resistance falls rapidly with decreasing current and is zero below the critical current. At $T>T_g$, the sample is in the vortex liquid state, the V-I curves exhibit ohmic behavior and the resistance remains constant even at small current I. Previous theoretical and experimental studies show that z=4~7 and v=1~2 are the reasonable values for VG phase transition [27]. In order to get a nice scaling performance, we have D=2, v=1.9, z=4.2.

Since magnetic field and temperature exhibit analogous effects in suppressing superconductivity and generating quasi-particles in conventional superconductors, I-V relations under different magnetic fields, which are shown in Fig. 3(c), are qualitatively similar to the I-V curves at different temperatures in Fig.3(a). The difference is that these V-I curves the occurrence of $V\propto I^3$ [30]. Below $T_{BKT}$, the vortex/antivortex pairs generated by topological excitations will be bound leading to a zero linear resistivity. At $T_{BKT}$, it has $V\propto I^3$, which is a sign of the BKT transition. The observed VG transition agrees with the previous reports that that InN is a type-II superconductor [12]. The quasi-2D VG character indicates that sample A's thickness ~900nm is comparable or smaller than the $\xi_0$ and the corresponding $\xi_g$.

*3.3. The non-zero resistance state at the limit of $T\to 0$.*

Different from sample A, sample B is sensitive to acid etching. Fig.4 (a) displays the R-T curves of the etched sample B. Compared with the as-grown sample B, it finds the superconducting transition is much broadened after etching. And the etched sample B's resistance is not zero even at the lowest temperature T=30mK. R-H measurements in Fig. 4(b) shows that etched sample B's "critical" magnetic field is also significantly reduced, which may be due to the reduced $T_c$.

Fig.4(d) is the I-V curves of etched sample B under different temperature. With increased temperature, the I-V curves show an evolution from convex curvature to linear dependence, which indicates the vortex glass state doesn't exist. The trend toward BKT transition with $V \propto I^3$ is observed under reduced temperature. The resistance goes to a finite value as the current I approaches zero, i.e. $V \propto I$ at $I \to 0$. In Fig.4 (e), we try to scale the I-V curves in Fig.4(d), even the attempt fails. The fitting parameters cannot be adjusted to the reasonable range, e.g. z=1.17 being away from the z=4~7 range.

The failed scaling is mainly due to the appearance of a non-zero resistance state (NZRS) at the limit of $T \to 0$[33]. As shown in Fig.4(c). The NZRS is not a vortex liquid state with $T_g$=0K, because the deuced $I \to 0$ linear resistance $R_{lin}$ in Fig.3(a) is not linear to log(T), as required by Eq.(5) [30]. This NZRS is also inconsistent with other mechanism that can suppress the superconductivity, like phase slip, in which the phase of the superconducting condensate wave-function jumps irreversibly, leading to the non-zero resistance. These phase slips represent activation processes that can be triggered either thermally (i.e. thermally activated phase slips, TAPS) or through quantum tunneling (quantum phase slip, QPS) [33]. However, TAPS requires a temperature dependence of resistance as[33]:

$$R(T) \propto \exp(-U/k_B T) \quad (7)$$

where U is the activation energy. But NZRS doesn't follow such relation at $T \to 0$, as shown in Fig.4(c, inset). On the other hand, QPS explanation requires the sample's sheet resistance $R_s$ in normal state is larger than the quantum of resistance

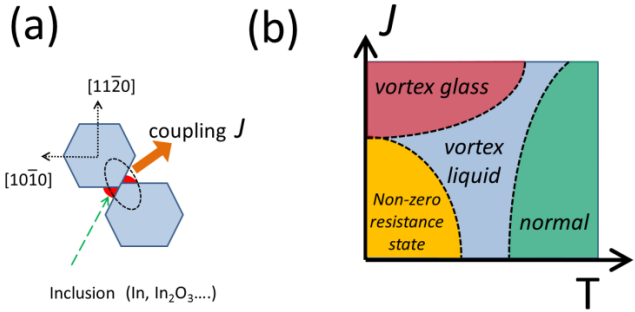

Fig.5. (a) Schematics of the inter-grain superconducting coupling in InN(0001) granular film. The presence of inter-grain inclusion can enhance the inter-grain coupling J. If such inclusions are removed, the grain boundary can only serve as a weak link with small J. (b) Schematic phase diagram of InN superconductor with the introduction of vortex (under a small magnetic field or with the persistent current's self-field), as a function of J and temperature T.

$R_Q$=h/4$e^2$=6.45kΩ[34], while our InN sample's $R_s$ is too small(~10Ω).

According to the phase transition theory, the scaling behavior can be attributed to the existence of a characteristic length which diverges at the transition temperature [10], like Eq.(4). Therefore, we propose that there should be a special size which limits the expansion of vortex-glass correlation length $\xi_g$ in the etched sample B even at $T \to 0$K. We believe the InN grain size can be such limit. On the other hand, since NZRS is not observed in the as-grown sample B, as shown in Fig.4(a), the In/$In_2O_3$ inclusion seems to "suppress" the NZRS.

Fig.5(a) gives our theoretical model regarding NZRS. In this model, we suggest that several factors are highlighted in InN superconductivity. Firstly, the grain size L makes difference. If L is comparable to or smaller than coherence length $\xi_0$ (e.g. ~700nm above), the magnitude fluctuations of superconducting order parameter will destroy the superconductivity [15]. Consequently, such grain is no longer superconducting. This situation will not change even if the InN is shaped into a [0001] orientated thin wire (i.e. in a quasi-one dimensional with the diameter <$\xi_0$ or ~$\xi_0$) as the growth going on [25]. Secondly, the inter-grain coupling can recover the superconductivity in InN. In this case, the diffusion of electron pairs from the superconductor grain into the normal material (known as the proximity effect) then into the neighboring grain gives rise to global superconductivity.

Fig.5(b) gives the schematic phase diagram describing InN superconductivity with the presence of vortex, as a function of inter-grain coupling J and temperature T. Since superconductivity on individual islands is fragile by its small size L<$\xi_0$ or L~$\xi_0$, non-superconducting T=0 states is present as the NZRS. The appearence of NZRS signifies the existence of mesoscopic energy scale on each grain (e.g. an effective "charging energy" or the electronic level spacing on a grain etc.) [15,32]. Therefore, a minimum inter-island coupling $J_m$ must be overcome for superconductivity to be attained [15].

*3.4. Discussion on the sample-dependent superconductivity in InN.*

With this phase diagram, we can consistently understand the various sample-dependent superconductivity results in InN based on the grain size and inter-grain coupling J. For those small grain sized InN film (e.g. sample B), since the superconductivity in each grain is fragile or not present, the sample's global superconductivity heavily relies on the inter-gain coupling which is realized by the inter-grain inclusion In/$In_2O_3$. The previous reports also clearly demonstrated that indium metal prefers to reside on the (11-20) planes of InN, which is the inter-grain interface of InN. Acid etching can then remove such inclusions and reduce the inter-gain coupling J. Consequently, a broadened transition does result not only simply from a spread in $T_c$ of the individual grains, but also from an inhomogeneous distribution of J. NZRS will also appear due to J<$J_m$. For large grain sized InN film (e.g. sample A), its individual grain has robust superconductivity and its inter-grain coupling is already strong (because of the large area of inter-grain touching interface). Therefore, acid etching has no obvious influences. The morphology observed in Fig.1(c,d) and previous reports [21] agree with the grain-size attribution above.

On the roles of In/$In_2O_3$ in InN superconductivity, we believe that they can enhance the superconductivity by filling the gaps between InN grains. However, if In/$In_2O_3$ alone want to establish the global superconductivity, they must: (a) continuously distribute from the source to the drain and get exposed there; (b) with a grain size >~30nm for In/In-oxide composite film [35] or >4nm for pure metal In[15]. However, item (a) and (b) will make In/$In_2O_3$ easy to be etched away by acid, so removing the superconductivity. Kadir et al. claimed $In_2O_3$ alone contributes to the superconductivity in InN via their thermal annealing experiment [17]. However, their annealing temperature is too higher (530-650°C) than the growth temperature. The grain size and inter-grain coupling had been certainly modified there.

## V. SUMMARY

We have measured I-V curves in HCl acid etched InN films at various magnetic fields and temperatures. We proved that InN superconductivity can be chemical stable against acid treatment, or sample-dependent. The standard vortex glass, vortex liquid phase in type II superconductor and a special non-zero resistance phase (which is typical in a weakly coupled superconducting islands system) are observed. The competing, transition among these phases are understood in a framework considering grain size and inter-grain coupling, which are the two key factors in InN superconductivity.

**Author Contributions.** T.K. initiated and supervised this research. Z.S. performed the measurements. Z.S. and T.K. analyzed the data. T.K. wrote the manuscript. P.C. provided the InN samples for the early stage of this research. A.Y. provided the InN samples used in this research. The requests for materials should be addressed to T.K..

**Acknowledgements.** The work was supported by the National Natural Science Foundation of China (NSFC) 11204334, 61475178, 61574150, 61376015, 91321311, Shanghai Science and Technology Foundation 14JC1406600. T. K. specially thanks the "Hundred Talent program" of the Chinese Academy of Sciences for the fund to establish the low-temperature transport lab.